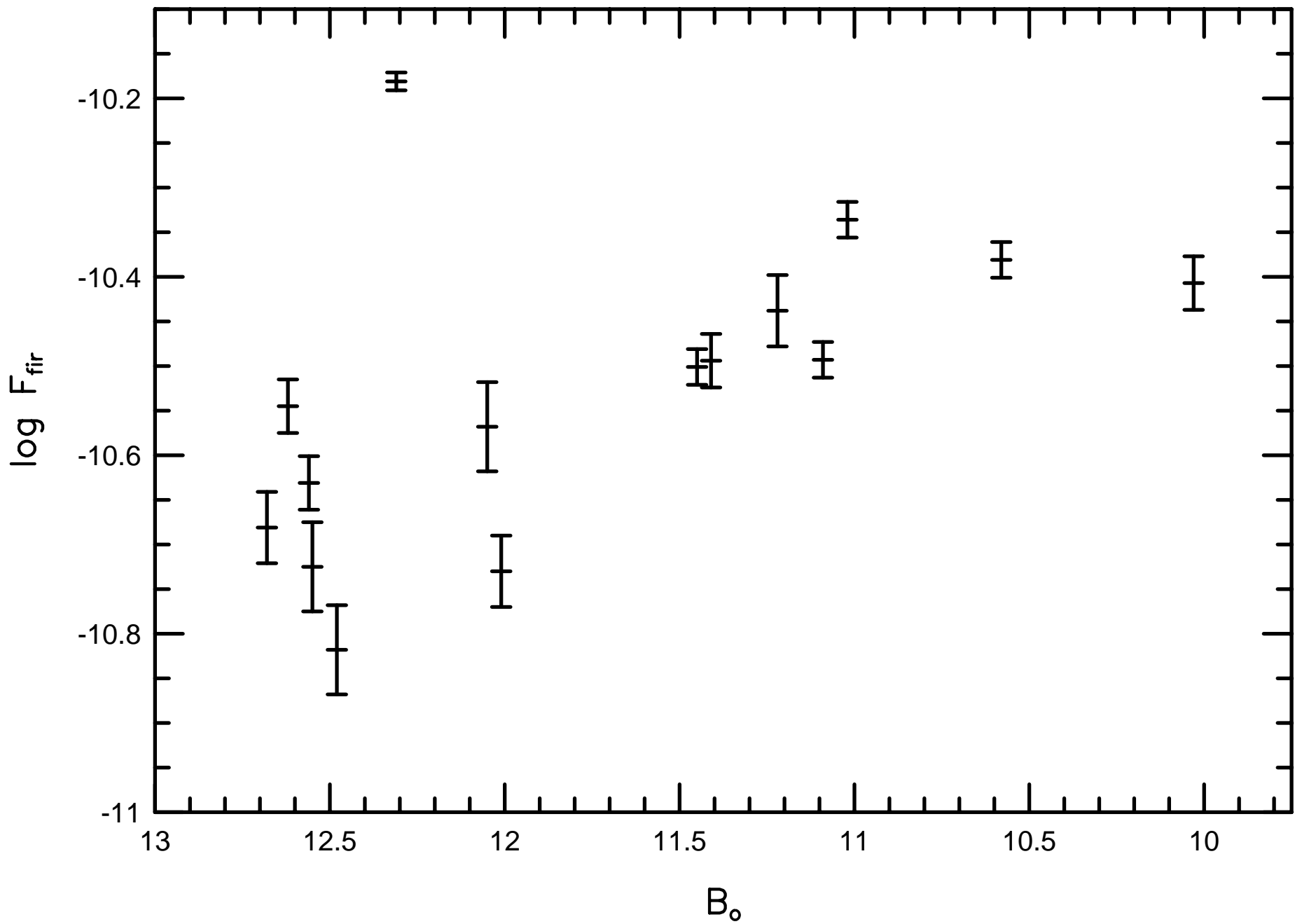

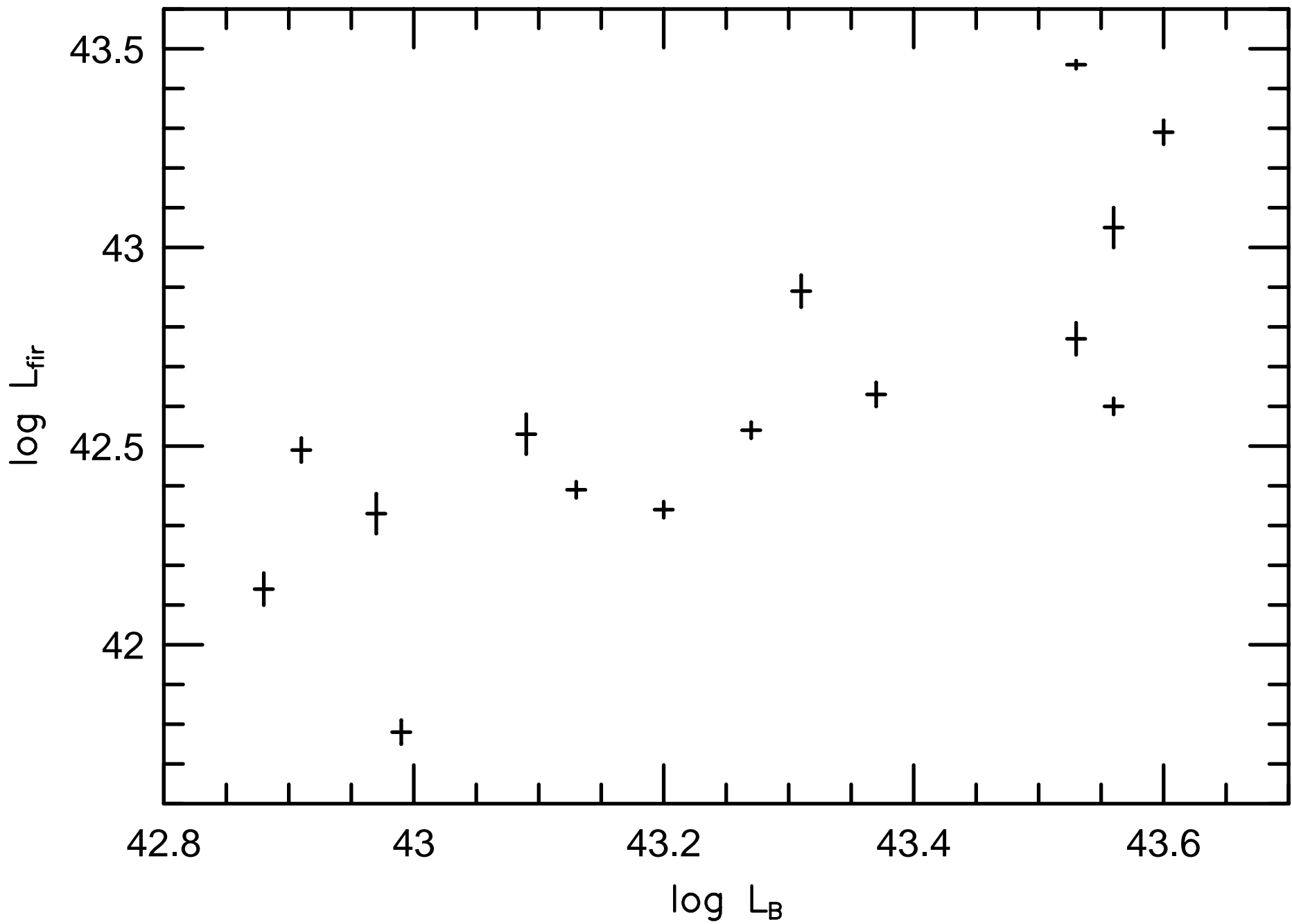

# Far-Infrared Emission From E and E/S0 Galaxies

Joel N. Bregman[1], Brian A. Snider[1], Laura Grego[1,2], and Caroline V. Cox[1,3]

1. Department of Astronomy, University of Michigan, Ann Arbor, MI 48109-1090
2. Astronomy Department, California Institute of Technology, Pasadena, CA 91125
3. Department of Astronomy, University of Virginia, Charlottesville, VA 22903-0818

**Abstract**
Early-type galaxies are filled with hot X-ray emitting gas, but the study of the less plentiful cold gaseous component has been more challenging. Studies of cold material through *IRAS* 60µm and 100µm observations indicated that half of ordinary E and E/S0 galaxies were detected above the 3σ level, indicating that cold gas is common, although no correlation was found between the optical and far-infrared fluxes. Most detections were near the instrumental threshold, and given an improved understanding of detection confidence, we reconsider the 60µm and 100µm detection rate. After excluding active galactic nuclei, peculiar systems, and background contamination, only 15 non-peculiar E and E/S0 galaxies from the RSA catalog are detected above the 98% confidence level, about 12% of the sample. An unusually high percentage of these 15 galaxies possess cold gas (HI, CO) and optical emission lines (Hα), supporting the presence of gas cooler than $10^4$ K. The 60µm to 100µm flux ratios imply a median dust temperature for the sample of 30 K, with a range of 23-38 K.

These detections define the upper envelope of the optical to far-infrared relationship, $F_{fir} \propto F_B^{0.24\pm0.08}$, showing that optically bright objects are also brighter in the infrared, although with considerable dispersion. A luminosity correlation is present with $L_{fir} \propto L_B^{1.65\pm0.28}$, but the dust temperature is uncorrelated with luminosity. The dust masses inferred from the far-infrared measurements are an order of magnitude greater than those from extinction observations, except for the recent merger candidate NGC 4125, where they are equal. We suggest that the ratio of the far-infrared dust mass to the extinction dust mass may be an indicator of the time since the last spiral-spiral merger.

These results are compared to the model in which most of the dust comes from stellar mass loss and the heating is primarily by stellar photons. Models that contain large dust grains composed of amorphous carbon plus silicates come close to reproducing the typical 60µm to 100µm flux ratios, the far-infrared luminosity, and the $L_{fir}$ - $L_B$ relationship.

## 1. Introduction

One of the many surprises from the *Infrared Astronomical Satellite* mission (*IRAS*) was that early-type galaxies were sources of emission in the far-infrared (60µm and 100µm), implying the presence of cold gas in galaxies once thought to be entirely devoid of interstellar material (Jura et al 1987). This discovery raised several observational issues, such as the detection frequency and associated fluxes, and physical issues such as the origin of the dust and its heat source.

The establishment of the observational parameters has proven to be challenging because the detection of early-type galaxies occurs near the threshold of the instrument. Consequently, accurate detection rates can depend sensitively on the nature of the properties of the underlying Galactic cirrus emission, among other issues. In a comprehensive work, Knapp et al. (1989) gave *IRAS* fluxes for a large number of galaxies of many types, where they found that the E and S0 galaxies were detected at or above the $3\sigma$ level at 60µm and 100µm in about half of the systems. They found that the ability to detect far-infrared emission in elliptical galaxies was independent of apparent magnitude or redshift of the system (also found by Goudfrooij and de Jong 1995), which would imply an extremely broad infrared luminosity distribution for galaxies of a given optical brightness. These data are essential in understanding the origin and evolution of dust in these systems (e.g., Tsai and Mathews 1996). For example, the lack of correlation between the optical and far-infrared luminosity has been interpreted as evidence for an infall origin of dust and gas (Goudfrooij 1994a,b; see also Forbes 1991), as opposed to the production of dust material by stars.

Following the work of Knapp et al. (1989), a deeper understanding has developed of *IRAS* fluxes at low levels of intensity so it becomes sensible to revisit the issue of far-infrared elliptical galaxy fluxes. Moshir et al. (1992) found that the noise properties of 60µm and 100µm *IRAS* fluxes do not obey a Gaussian distribution, which will affect the confidence of source fluxes, especially near the detection threshold. This was evident to various investigators, such as Haynes et al. (1990), who assigned qualitative measures of the reality of detection. A more quantitative analysis of low flux level detections was undertaken by Cox, Bregman, and Schombert (1995), who examined the fluxes returned from 60µm and 100µm strip scans (ADDSCAN, now SCANPI) in 207 randomly chosen locations. They found that if one uses the rms provided by the standard *IRAS* processing (ADDSCAN or SCANPI), the spurious detection rate for a $3\sigma$ threshold is 16% at 60µm and 31% at 100µm, whereas the false detection rate should have been 0.13% (for a normal distribution, a 16% detection rate corresponds to $+1.0\sigma$ and a 31% rate to $+0.24\sigma$). The analysis of the control fields reveal that in order to have a detection at a confidence levels of 98% using the standard IPAC data products, one should employ a threshold of $4\sigma$ at 60µm and $4.5\sigma$ at 100µm, while a 99.98% confidence level is achieved with a threshold of $6\sigma$ at 60µm or 100µm (Cox, Bregman, and Schombert 1995).

In this work, we reexamine the detection rate and fluxes for E and E/S0 galaxies using our improved knowledge of the statistics of low light level *IRAS* sources. It is not the intent of this work to demonstrate that few E and S0 galaxies are far-infrared emitters, since all must emit such radiation at some level. Our goal is to define a moderate number of early-type galaxies for which good fluxes exist and then to use this sample to try to understand the origin of far-infrared emission. Eventually, it is hoped that we can distinguish between the far-infrared emission by stellar mass loss spread through the galaxy from that by a cold disk of material.

## 2. Sample Selection and Detection Rate

From the catalog of early-type RSA galaxies, we selected the galaxies of type E and E/S0 (code types 10 and 30 in Roberts et al. 1991), avoiding any galaxies listed as peculiar. In the following analysis, we consider the 153 E and E/S0 galaxies together as one group (124 E galaxies and 29 E/S0 galaxies). Of this sample, six were not observed with *IRAS*, reducing the sample size to 147.

For these 147 E and E/S0 galaxies, Knapp et al. (1989) list 74 galaxies that were detected by *IRAS* at or above the $3\sigma$ level in the 60µm and/or the 100µm band and 73 that fall below this detection threshold. We have reextracted the fluxes for each detected galaxy but not from the sources with upper limits, which are taken to be correct. For data extraction, the IPAC routine ADDSCAN was used (now SCANPI), which extracts flux from strip-scans. Also, 60µm and or the 100µm images were used (FRESCO), which has lower point source sensitivity but is valuable in assessing contamination issue. Finally, an independent extraction procedure of the strip scans was employed as a cross-check of the ADDSCAN results and this is described in Cox, Bregman, and Schombert (1995). We examined this emission from each of these detected galaxies and apply more stringent criteria in order to assure that the emission is thermal and originates in the galaxies at above the 98% confidence level. Consequently, we rule out known or likely AGN sources, such as a galaxy with a strong radio continuum, defined here as having a flux density above 0.5 Jy at 1.4 GHz (these two criteria eliminate NGC 1052, NGC 4374, NGC 4486, NGC 5090, NGC 741, IC 4296, and IC 1459). Although the detected far-infrared emission may not be related to central AGN, we can not with certainty distinguish between far-infrared sources that are powered by AGNs and those that are not; we have taken the conservative approach of eliminating such targets from the sample. We eliminated detections where there was a difference of more than 1' between the optical center (from NED) and the 60µm or the 100µm emission. This is because the positional uncertainty of *IRAS* emission is significantly less than 1', and an offset of 1' will place the emission beyond $D_{25}$ for many of these galaxies (this eliminates the galaxies NGC 5638, NGC 4168, NGC 821). For several sources, we determined that there was a foreground or background object in the scan line that might have contaminated the scan, leading to a false detection (e.g., NGC 1453, NGC 1600, NGC 2986, NGC 3268, NGC 6876, NGC 4168, NGC 2749, NGC 4742). Some sources were eliminated because they are located in regions where the background was rapidly changing so that the extraction of flux was suspect (e.g., NGC 1453, NGC 3136, NGC 3557, NGC 4645, NGC 5090, NGC 7041). This procedure removes 21 of 74 galaxies and places three into the null detection group (the three where the emission is offset by more than 1') and 18 into an ambiguous group where we cannot determine whether warm dust is present at low flux levels (due to AGNs, nearby spiral galaxies, etc.). We remove these 18 from further consideration, leaving us with 129 galaxies in the sample. We found that of the remaining 53 galaxies for which $3\sigma$ or greater detections were reported, only 15 objects that are detected by us at or above the 98% confidence level (e.g., above $4\sigma$ at 60µm and $4.5\sigma$ at 100µm; Table 1, 2), and seven additional galaxies have detections in the 90-98% confidence level range.  Therefore, the detection rate for these galaxies is 12-17%, depending upon whether one uses the 15 excellent detections (15/129) or the seven likely detections as well (22/129), which is much lower than the 48% of Knapp et al. (1989) or the 38% of Roberts et al. (1991; they used the Knapp et al. values, but with a 3.5σ threshold).

One issue that we have considered is whether the statistics have been biased by excluding "peculiar" galaxies, since this designation may have arisen due to the presence of dust extinction. Consequently, we examined in detail the 14 E and E/S0 galaxy listed as peculiar in the RSA catalog and performed an analysis identical to that for the non-peculiar galaxies.  We eliminated the nearby dwarf galaxies around M31 since dwarfs were not considered in the sample (NGC 185 and NGC 205) and the famous AGN/cooling flow galaxy in the center of Perseus,  NGC 1275. Of the remaining galaxies, 8 are peculiar due to galaxy-galaxy interactions (i.e., they have tails, shells, ripples; IC 3370, NGC 2672, NGC 2832,NGC 4782, NGC 4783, NGC 5576, NGC 6776, and NGC6854), which is evident from studying the images in the new Carnegie Atlas (1994) or from reading the comments associated with that compilation and other investigations.  Two galaxies are peculiar due to dust lanes or patches (NGC 4476 and NGC 5796; one has been reclassified into an $S0_3$ in the new Carnegie Atlas, NGC 4476), and one is no longer listed as peculiar in the Carnegie Atlas (NGC 4760).  Evidently, it is the presence of mild galaxy-galaxy interactions that generally lead to the designation as "peculiar" rather than the presence of dust.

We analyzed the peculiar galaxies identically to the non-peculiar sample objects in order to determine the fraction of the peculiar E and E/S0 galaxies that are IRAS sources.  After excluding the AGN and the strong radio source (2 objects),  we examined every published detection (by Knapp et al. 1989) at both 60µm and 100µm and found that one detection was false due to a contaminating galaxy along the scan line (NGC 2832; this is the central galaxy of Abell 779 so it is not surprising to find that the detection was confused with another galaxy; also, there is a significant offset in the position of the detection).  One detection is well below our threshold and extremely unlikely (NGC 2672), another is an excellent detection and would be in our first category (>98% confidence category; IC 3370), while one galaxy is a likely detection and would fall in the 90-98% confidence detection bin (NGC 6776; the detection at 60µm might be real, but the "source" at 100µm is not at the same location as the 60µm source and is beyond the half-light radius of the galaxy).  To summarize, of the 9 remaining objects in the sample, 7 are non-detections, 1 is a definite detection, and 1 is a possible detection.  Based on the statistics of the non-peculiar galaxies, we would have expected 1.1 definite detections and 0.5 likely detections . Consequently, the detection rates for peculiar and non-peculiar galaxies are similar.  The subsequent analysis deals with the non-peculiar galaxies, but the results would not change had the peculiar systems been included.

## 3. Analysis of the Detections

The detected objects suggest a relationship between detectable far-infrared emission and the presence of cold gas.  For the sample of 15 excellent detections (the sample most commonly employed in the following analysis), 27% (4/15) have detected HI or CO (NGC 2974, NGC 4278, NGC 5353, and NGC 7196), while in the total sample of E and E/S0 galaxies from the RSA catalog, only 8% have detected HI or CO (Roberts et al. 1991).  For the remaining 11/15 galaxies without known cold gas, five have never been observed at 21 cm for HI (NGC 4786, NGC 4936, NGC 5357, NGC 6851, and NGC 7097) and two have 21 cm upper limits, but the observations were not particularly sensitive (NGC 4125, and NGC 4697), so HI may be present in several additional of these galaxies. Furthermore, although NGC 3156 has not been detected in cold gas, it contains a young stellar population (Worthey, private communication), a very unusual situation that implies the presence of cold gas in the recent past. There are only 3/15 far-infrared detected galaxies with good upper limits to their cold gas content (NGC 2693, NGC 5322, and NGC 6868). All three are among the intrinsically most optically luminous galaxies in the sample, although we do not have an insight as to why this might affect their far-infrared emission.

When searching for a connection between the far-infrared emission and the presence of cold gas, it must be noted that for present levels of sensitivity, for a normal gas to dust ratio (100-200), it is easier to detect 60µm or/and 100µm *IRAS* emission than it is to detect the 21 cm HI line or the CO(1-0) transition (after Roberts et al. 1991). Nevertheless, it is suggestive that an unusual number of far-infrared detected galaxies have evidence for cold gas and it would be worthwhile to have sensitive HI and CO measurements for galaxies not yet adequately observed.

The far-infrared detected ellipticals are further distinguished in that they are detected in optical emission line observations in an unusually high fraction of cases. Of the 56 elliptical galaxies studied by Goudfrooij et al. (1994a) for Hα line emission, 31 were detected (55%). Of our sample of 15, five were observed by Goudfrooij, all of which were detected (an additional 2/3 sources in our extended list were detected). From all galaxies in the literature with unambiguous results (compiled in Roberts et al. 1991), 11/14 ellipticals show line emission (78%), while only 29% of all RSA ellipticals were reported with positive detections in that compilation.

In order to compare the far-infrared properties to optical fluxes and luminosities, we adopt the total apparent magnitudes from Faber et al. (1989) as well as their velocity distances. In the few cases that these were not available, we used the RSA magnitudes and their recession velocities corrected for Galactic rotation and Virgocentric infall. We use the usual "bolometric" far-infrared brightness

$$F_{fir} = 1.24 \times 10^{-14} (2.58 F_{60\mu m} + F_{100\mu m}) \text{ erg cm}^{-2} \text{ s}^{-1},$$

where the 60µm and 100µm flux densities are expressed in mJy. We find a clear correlation between the optical and far-infrared brightness of the 15 detected galaxies (Fig. 1). The obvious outlier is NGC 7196, the only galaxy in the sample with detected CO emission (Huchtmeier and Tammann 1992) and with $F_B/F_{fir} \approx 1$ (the typical value for this sample is 5); CO emission from E and E/S0 galaxies is extremely rare. The correlation between $F_{fir}$ and $B_o$ is present because there are no optically faint sources ($B_o > 11.9$) that are brighter than $\log F_{fir} = -10.55$ (with the exception of NGC 7196), while all 7 sources brighter than $B_o = 11.5$ have $\log F_{fir} \geq -10.5$. There are many sources with fainter $\log F_{fir}$ at some given $B_o$ (the galaxies undetected by *IRAS*), but none brighter. This correlation is hardly surprising in that it indicates that the optically brighter galaxies are also brighter in the infrared.

It must be recognized that this correlation does not yield the true relationship between the far-infrared and optical brightness, since we are able to measure the far-infrared brightness for only a small fraction of galaxies, the vast majority of sources being upper limits. The correlation given in Fig. 1 is the upper envelope of the distribution. The slope of the correlation between $\log F_{fir}$ and $\log F_B$ of this upper envelope is $0.24 \pm 0.08$ (i.e., $d\log F_{fir}/d\log F_B$), using the least squares bisector method (Feigelson and Babu 1992). This value of $0.24 \pm 0.08$ is significantly different from a linear relationship (slope of 1.0), indicating that the intrinsically optically bright galaxies are not proportionally bright at 60µm and 100µm, at least when examining the top of the detection envelope.

We examined the optical and far-infrared luminosity relationship ($L_{fir}$ vs $L_B$; Fig. 2), although there is always the pitfall that for a flux-limited sample, the presence of distance squared in the two luminosities can lead to a linear correlation when none exists. To assess this effect, we performed Monte-Carlo simulations whereby the fluxes and distances were scrambled before calculating luminosities. As anticipated, the simulations lead to a slope near unity (1.04 $\pm 0.13$) with substantial scatter. When we measure the correlation coefficient of the Monte-Carlo simulations, it is $R^2 = 0.35 \pm 0.076$. From the data, we find a correlation coefficient of $R^2 = 0.62$, which is a stronger correlation compared to the simulations at the 99.5% confidence level. Also, the slope of the fit is steeper than unity ($d\log L_{fir}/d\log L_B = 1.65 \pm 0.28$, using the bisector method). The outlier galaxy in the flux-flux plot, NGC 7196, is no longer a deviant point, as it is one of the most optically luminous objects in the sample (the slope does not change significantly if it is removed from the sample). We examined whether upper limits could add to the analysis and lead to an improved $\log L_{fir}$ - $\log L_B$ relationship, but found that the upper limits were not distributed randomly in $L_{fir}$ - $L_B$ space, an underlying assumption of statistical methods with censored data (Isobe, Feigelson, and Nelson 1986).

The temperature of the dust is given approximately by $T_d \approx 49 (F_{60\mu m}/F_{100\mu m})^{0.4}$ (Young et al. 1989), although the precise value of the temperature depends upon the composition and size distribution of the dust grains. In our sample of detected sources, the median value of $F_{100\mu m}/F_{60\mu m}$ is 3.4, so the median dust temperature is 30 K, although there is a clear range in the mean dust temperature for each galaxy, ranging from $22.9\pm1.9$ for NGC 6851 to $37.5\pm1.4$ for NGC 4125. These temperatures are typical of dust in spiral galaxies with low or modest star formation rates (e.g., Soifer, Houck, and Neugebauer 1987). There is no correlation between the temperatures and the optical or infrared luminosity of the galaxies or whether they have been detected in HI.

The dust masses can be calculated by combining the inferred temperature information with the far-infrared flux level (Young et al. 1989)

$$M_{dust} = 0.00478 \, F_{100\mu m} \, d^2 \, \{\exp(2.94(F_{100\mu m}/F_{60\mu m})^{0.4})-1\} \, M_\odot \, ,$$

where $F_{100\mu m}$ and $F_{60\mu m}$ are given in mJy and the distance to the galaxy, d, is in Mpc. For the few galaxies where Goudfrooij et al. (1994a,b) has obtained dust masses from extinction observations, the far-infrared dust mass is generally an order of magnitude greater, as was noted by Goudfrooij and de Jong (1995) and others. He suggests that the difference in the dust masses can be understood if most of the dust is smoothly distributed throughout the galaxy. However, we note that the galaxy NGC 4125 is distinguished in that the far-infrared dust mass is the same as that derived from optical extinction measurement, to within the uncertainties (about 30%; this object is discussed further below).

**4. Discussion and Conclusions**

We obtained a sample of high-quality detections of far-infrared emission from RSA E and E/S0 galaxies that are not listed as peculiar or known to be AGNs. The resulting sample of 15 excellent detections (along with 7 likely detections; 12-17% of the RSA E and E/S0 galaxies) have a $F_{100\mu m}$ to $F_{60\mu m}$ ratio that is similar to that found in spiral galaxies and is consistent with dust emission at a mean temperature near 30 K. These 15 galaxies have a much higher than normal incidence of optical emission line material (Hα) and cold gas (HI, CO), supporting the dust emission interpretation. There is a positive correlation between the optical apparent brightness and the far-infrared brightness (at 60µm and 100µm), which is expected but was not seen previously. This correlation, $F_{fir} \propto F_B^{0.24}$ reflects the top of the detection envelope since 85-90% of the E and E/S0 galaxies are not securely detected. When viewed in luminosity space, $L_{fir} \propto L_B^{1.65}$, and this may be an important clue to the origin of the dust and the dust heating mechanism.

There are two likely origins for the dust in these elliptical galaxies: mass loss from stars in the galaxy, or capture of material from a moderate luminosity system (so that the metallicity is high enough to have produced dust). Although there are clear examples of accretion, such as in the case of NGC 4278 (Raimond et al. 1981), stellar mass loss must be an important phenomenon that adds dust to the typical sample galaxy at a rate of ~$10^{-3}$ $M_\odot$ yr$^{-1}$ (Faber and Gallagher 1976; Knapp, Gunn, and Wynn-Williams 1992), so the observed *IRAS* dust mass can be accumulated in ~$10^9$yr, ignoring dust destruction. Further support for a stellar origin of some dust comes from the discrepancy between the *IRAS* dust mass and the extinction dust mass, with the later being about an order of magnitude smaller. This discrepancy can be understood if most of the dust is extended, as would occur when stellar mass loss distributes dust throughout the galaxy (Goudfrooij and de Jong 1995).

The primary shortcoming with the above picture is that detailed models fail to predict the observed dust masses and dust temperatures (Goudfrooij and de Jong 1995; Tsai and Mathews 1995, 1996). In the most detailed calculations to date, Tsai and Mathews (1995, 1996) calculate the evolution of dust expelled from stars, taking into account sputtering by the hot ambient medium, grain size distribution, inflow, and heating by starlight, X-rays, and hot gas. For an initial dust size distribution given by Mathis, Rumpl, and Nordsieck (1977) and either a graphite and silicate dust mixture or an amorphous carbon and silicate mixture, the far-infrared luminosities are 0.5-0.8 dex too low, because sputtering reduces the dust mass, and the predicted ratio $F_{100\mu m}/F_{60\mu m}$ is 1-2 rather than the typical observed values of 3-4.

These problems might be resolved with a dust grain size distribution containing grains larger than the maximum size suggested for the Milky Way by Mathis, Rumpl, and Nordsieck (1977), as this would lengthen the sputtering time and increase the steady-state dust mass. When Tsai and Mathews (1996) increased their maximum grain size from 0.3 µm to 0.9 µm, the far-infrared luminosity increased as did $F_{100\mu m}/F_{60\mu m}$. For a grain mixture of amorphous carbon and silicates, Tsai and Mathews (1996; Model a, AC+) find $F_{100\mu m}/F_{60\mu m} = 3.4$, the average value we observe, and they find a bolometric far-infrared luminosity of $4.6 \times 10^{42}$ erg s$^{-1}$, which is in the range of observed values for the corresponding optical luminosity. However, this runs counter to the conclusions of Goudfrooij and de Jong (1995) who argue that the mean grain size is smaller than that in the Milky Way.

The far-infrared luminosity is predicted to increase as $L_B^{1.1}$ (Tsai and Mathews 1996),

which is a bit shallower than the $L_B^{1.65\pm0.28}$, but the difference may not be significant considering the limitations of the *IRAS* sample. To summarize, the model discussed by Tsai and Mathews (1995, 1996), with the modification of large dust grains, can reproduce some of the basic observational properties of our sample. A substantial improvement in defining the $L_B$ - $L_{fir}$ relationship will be realized with the new infrared data that should be forthcoming from the *Infrared Space Observatory*. Advances in the theoretical work will be aided by more detailed knowledge of the hot gas density distribution (from *ROSAT, ASCA,* and eventually *AXAF*), which bear upon the calculation of the sputtering rate.

The galaxy NGC 4125 is of special interest because it is the only object that has the same dust mass determined from extinction observations and from the $F_{100\mu m}$ and $F_{60\mu m}$ measurements. It also stands out as one of the only early-type galaxies identified by Schweizer and Seitzer (1992) as a recent likely merger of two spiral systems, based upon their fine structure parameter. Four other ellipticals suggested to be recent mergers are listed by Schweizer and Seitzer (1992), but only one has an extinction dust mass, NGC 3640 (log $M_d$ = 4.30). However, the upper limit to the *IRAS* dust mass for NGC 3640 is log $M_d$ < 5.51, so more sensitive far-infrared observations will be required to determine if the extinction and *IRAS* dust masses are equal. The merger picture may fit together with the equality of the extinction and *IRAS* dust masses. Following the merger of two spirals, nearly all of the dust would be in the remaining disk of cold gas and dust, which is detected both through far-infrared measurements and through extinction (dust lanes, disturbance of elliptical isophotes; Goudfrooij 1994a). As the stars in this new elliptical shed dust and gas, the amount of distributed dust and extinction grows, and although it is detected through its far-infrared emission, the extinction is distributed over the face of the galaxy and very difficult to detect; the usual extinction observation fails to measure the smoothly distributed dust and thereby underestimates the dust mass. Consequently, the ratio of the *IRAS* dust mass to the extinction dust mass would appear to grow after a merger and this may provide a new diagnostic of the time since a major merger.

We would like to acknowledge the substantial assistance provided by the *Infrared Processing and Analysis Center*, which is supported by NASA, both for providing the various *IRAS* data products and for the use of the *NASA Extragalactic Database*. We are thankful for the advice of W.G. Mathews, G. Helou, D. Levine, A. Wehrle, M. Roberts, D. Hogg, E. Feigelson, and E. Dwek, as well as a thoughtful and detailed report by the referee. Financial support for this project was provided by NASA grants NAGW-2135 and NAGW-4448.

# Figure Captions

Figure 1. The bolometric far-infrared flux vs the total optical apparent magnitude (at B) of the galaxy for the 15 high-quality detections. The outlier is NGC 7196, the only galaxy in the sample known to possess molecular gas through detection of the CO(1-0) line. Excluding NGC 7196, the slope of the relationship is $F_{fir} \propto F_B^{0.24\pm0.08}$.

Figure 2. The bolometric luminosity vs the B band optical luminosity for the 15 E and E/S0 galaxies detected with *IRAS*. NGC 7196 has a high optical luminosity so it does not stand out as an outlier (it has the highest $L_{fir}$). Using the bisector method, we find that $L_{fir} \propto L_B^{1.65\pm0.28}$.

**Table 1: Optical Properties Of Sample E and E/S0 Galaxies**

| Galaxy Name | Type | Bo | Rvel (km/s) | D (Mpc) | logLo (LB,sun) | logLo (erg/s) |
|---|---|---|---|---|---|---|
| IRAS detections above the 98% confidence level | | | | | | |
| NGC 2693 | E | 12.55 | 4917 | 70.2 | 10.87 | 43.56 |
| NGC 2974 | E | 11.45 | 2122 | 30.3 | 10.58 | 43.27 |
| NGC 3156 | E/S0 | 12.01 | 1740 | 24.9 | 10.18 | 42.88 |
| NGC 4125 | E/S0 | 10.58 | 1986 | 28.4 | 10.87 | 43.56 |
| NGC 4278 | E | 11.02 | 1470 | 21.0 | 10.43 | 43.13 |
| NGC 4697 | E | 10.03 | 794 | 11.3 | 10.29 | 42.99 |
| NGC 4786 | E | 12.62 | 5293 | 75.6 | 10.90 | 43.60 |
| NGC 4936 | E | 11.22 | 2571 | 36.7 | 10.83 | 43.53 |
| NGC 5322 | E | 11.09 | 1661 | 23.7 | 10.51 | 43.20 |
| NGC 5353 | E/S0 | 12.05 | 2255 | 32.2 | 10.39 | 43.09 |
| NGC 5357 | E | 12.68 | 3895 | 55.6 | 10.61 | 43.31 |
| NGC 6851 | E | 12.56 | 2328 | 33.3 | 10.21 | 42.91 |
| NGC 6868 | E/S0 | 11.41 | 2328 | 33.3 | 10.67 | 43.37 |
| NGC 7097 | E | 12.48 | 2400 | 34.3 | 10.27 | 42.97 |
| NGC 7196 | E/S0 | 12.31 | 4244 | 60.6 | 10.83 | 43.53 |
| IRAS detections in the 90-98% confidence range | | | | | | |
| NGC 1366 | E/S0 | 12.81 | 1422 | 20.3 | 9.68 | 42.38 |
| NGC 1395 | E | 10.94 | 1990 | 28.4 | 10.72 | 43.42 |
| NGC 3258 | E | 12.27 | 3488 | 49.8 | 10.68 | 43.38 |
| NGC 3377 | E | 11.13 | 857 | 12.2 | 9.92 | 42.61 |
| NGC 3872 | E | 12.59 | 3604 | 51.5 | 10.58 | 43.28 |
| NGC 3904 | E | 11.67 | 1583 | 22.6 | 10.23 | 42.93 |
| NGC 4636 | E/S0 | 10.2 | 1333 | 19.0 | 10.67 | 43.37 |

**Table 2: Far Infrared Properties Of Sample E and E/S0 Galaxies**

| Galaxy Name | logLo (erg/s) | F60um (mJy) | err (mJy) | F100um (mJy) | err (mJy) | logFfir (erg/s/cm2) | logLfir (erg/s) | log err (erg/s) | Tdust (K) | err (K) | logMd Mo | log err | Comments |
|---|---|---|---|---|---|---|---|---|---|---|---|---|---|
| IRAS detections above the 98% confidence level | | | | | | | | | | | | | |
| NGC 2693 | 43.56 | 230 | 49 | 900 | 117 | -10.73 | 43.05 | 0.05 | 28.4 | 3.1 | 6.53 | 0.20 | no HI, Halpha |
| NGC 2974 | 43.27 | 420 | 33 | 1420 | 80 | -10.50 | 42.54 | 0.02 | 30.1 | 1.1 | 5.87 | 0.09 | 12, 25 um source, HI, Halpha detected |
| NGC 3156 | 42.88 | 220 | 37 | 910 | 95 | -10.73 | 42.14 | 0.04 | 27.8 | 3.0 | 5.68 | 0.17 | young stellar population; no HI, Halpha |
| NGC 4125 | 43.56 | 730 | 44 | 1420 | 117 | -10.38 | 42.60 | 0.02 | 37.5 | 1.4 | 5.39 | 0.09 | poor HI rms; strong Halpha; merger |
| NGC 4278 | 43.13 | 660 | 49 | 1960 | 111 | -10.34 | 42.39 | 0.02 | 31.7 | 1.3 | 5.58 | 0.08 | HI detected; strong Halpha |
| NGC 4697 | 42.99 | 590 | 53 | 1590 | 156 | -10.41 | 41.78 | 0.03 | 33.0 | 2.1 | 4.88 | 0.12 | 12, 25 um source; poor HI rms; Halpha |
| NGC 4786 | 43.60 | 370 | 36 | 1310 | 126 | -10.54 | 43.29 | 0.03 | 29.6 | 2.1 | 6.67 | 0.13 | no HI obs; Halpha emission |
| NGC 4936 | 43.53 | 410 | 56 | 1840 | 204 | -10.44 | 42.77 | 0.04 | 26.9 | 2.6 | 6.40 | 0.16 | no HI obs; Halpha emission |
| NGC 5322 | 43.20 | 550 | 26 | 1130 | 70 | -10.49 | 42.34 | 0.02 | 36.7 | 1.3 | 5.18 | 0.07 | no HI; Halpha emission |
| NGC 5353 | 43.09 | 340 | 88 | 1270 | 111 | -10.57 | 42.53 | 0.05 | 28.9 | 3.8 | 5.96 | 0.20 | HI present; no Halpha emission |
| NGC 5357 | 43.31 | 350 | 35 | 750 | 112 | -10.68 | 42.89 | 0.04 | 36.1 | 2.8 | 5.77 | 0.15 | no HI obs; no Halpha emission |
| NGC 6851 | 42.91 | 200 | 40 | 1340 | 76 | -10.63 | 42.49 | 0.03 | 22.9 | 1.9 | 6.58 | 0.19 | no HI obs; Halpha emission |
| NGC 6868 | 43.37 | 560 | 45 | 1100 | 160 | -10.49 | 42.63 | 0.03 | 37.4 | 2.3 | 5.43 | 0.14 | likely 12um; no HI; Halpha; Abell S0851 |
| NGC 7097 | 42.97 | 200 | 27 | 690 | 109 | -10.82 | 42.33 | 0.05 | 29.9 | 2.4 | 5.68 | 0.19 | no HI obs; Halpha emission |
| NGC 7196 | 43.53 | 900 | 34 | 2910 | 114 | -10.18 | 43.46 | 0.01 | 30.6 | 0.8 | 6.75 | 0.05 | likely 12, 25um; CO; no HI obs |
| IRAS detections in the 90-98% confidence range | | | | | | | | | | | | | |
| NGC 1366 | 42.38 | 70 | 35 | 590 | 75 | -11.01 | 41.68 | 0.07 | 20.9 | 3.5 | 6.06 | 0.39 | poor HI limit; Halpha emission |
| NGC 1395 | 43.42 | 100 | 25 | 430 | 79 | -11.06 | 41.92 | 0.06 | 27.3 | 3.9 | 5.51 | 0.25 | 12, 25 um source; no HI obs; Halpha |
| NGC 3258 | 43.38 | 250 | 37 | 1420 | 270 | -10.58 | 42.89 | 0.06 | 24.5 | 3.2 | 6.78 | 0.24 | likely 12um source; no HI; Halpha |
| NGC 3377 | 42.61 | 140 | 47 | 850 | 124 | -10.82 | 41.44 | 0.06 | 23.8 | 3.2 | 5.41 | 0.29 | no HI; weak Halpha |
| NGC 3872 | 43.28 | 120 | 55 | 750 | 106 | -10.87 | 42.63 | 0.07 | 23.5 | 5.0 | 6.63 | 0.35 | no HI; no Halpha |
| NGC 3904 | 42.93 | 370 | 48 | 540 | 126 | -10.73 | 42.06 | 0.05 | 42.1 | 6.6 | 4.59 | 0.20 | no HI; no Halpha |
| NGC 4636 | 43.37 | 300 | 34 | 500 | 164 | -10.79 | 41.84 | 0.06 | 39.9 | 8.1 | 4.49 | 0.26 | no HI; Halpha emission |